\documentclass[lettersize,journal]{IEEEtran}
\usepackage{amsmath,amsfonts}
\usepackage{algorithmic}
\usepackage{algorithm}
\usepackage{array}
\usepackage[caption=false,font=normalsize,labelfont=sf,textfont=sf]{subfig}
\usepackage{textcomp}
\usepackage{stfloats}
\usepackage{url}
\usepackage{verbatim}
\usepackage{graphicx}
\usepackage{cite}
\usepackage{bm}
\usepackage{algorithm}
\usepackage{algorithmic}
\usepackage{booktabs}
\usepackage{enumitem}
\usepackage[hidelinks]{hyperref}
\usepackage{tikz}
\usepackage{xcolor}
\usepackage{xspace}
\newcommand{\fig}[1]{{Fig.~#1}\xspace} 

\newcommand{\sect}[1]{{\S#1}\xspace} 
\newcommand{\tab}[1]{{Table~#1}\xspace} 
\newcommand{\head}[1]{{\noindent\textbf{#1.}\xspace}} 

\newcommand{\js}[1]{{\color{black}{#1}}} 
\newcommand{\mj}[1]{{\color{black}{#1}}} 

\newcommand{\mjr}[1]{{\color{black}{#1}}} 

\newcommand{\vpass}{$V_{pass}$\xspace} 
\newcommand{\vpassh}{$V_{passH}$\xspace} 
\newcommand{\vpassl}{$V_{passL}$\xspace} 
\newcommand{\bthr}{$RC_\text{MAX}$\xspace} 
\newcommand{\br}{$RC$\xspace} 
\newcommand{\wli}[1]{$\text{WL}_{#1}$\xspace}
\newcommand{\wlij}[2]{$\text{WL}_{#1}^{\text{#2}}$\xspace}
\newcommand{\blki}[1]{$\text{BLK}_{#1}$\xspace}
\newcommand{\ercmax}{$ERC_\text{MAX}$\xspace} 
\newcommand{\erc}{$ERC$\xspace} 

\newcommand{\maintech}{\textsc{Straw}\xspace} 
\newcommand{\ftl}{\textsc{Straw}FTL\xspace} 
\newcommand{\inum}[1]{(\textit{#1})}

\newcommand{\tr}{\texttt{tREAD}\xspace} 
\newcommand{\tprog}{\texttt{tPROG}\xspace} 
\newcommand{\tbers}{\texttt{tERASE}\xspace} 
\newcommand{\usec}{$\mu$s\xspace} 

\DeclareRobustCommand\blkcirc[1]{\tikz[baseline=(char.base)]{
           \node[shape=circle,draw=none,inner sep=0pt,fill=black, text=white] (char) {#1};}}
           
\newcommand{\ssdideal}{\textsf{\small{SSD$_{\text{ideal}}$}}\xspace}
\newcommand{\ssdreal}{\textsf{\small{SSD$_{\text{real}}$}}\xspace}

\newcommand{\block}{\textsf{\small{BLOCK}}\xspace}
\newcommand{\pt}{\textsf{\small{PAGETYPE}}\xspace}
\newcommand{\straw}{\textsf{\small{STRAW-SS}}\xspace}
\newcommand{\wl}{\textsf{\small{STRAW-WL}}\xspace}

\newcommand{\pa}{\textsf{\small{PA}}\xspace}
\newcommand{\pb}{\textsf{\small{PB}}\xspace}

\begin{document}

\title{\huge \maintech: A Stress-Aware WL-Based Read Reclaim Technique for High-Density NAND Flash-Based SSDs}

\author{
Myoungjun Chun, Jaeyong Lee, Inhyuk Choi, Jisung Park, Myungsuk Kim, and Jihong Kim
\thanks{Manuscript received August 24, 2024; accepted November 24, 2024. This work was supported by Samsung Electronics Co., Ltd (IO201207-07809-01) and IITP (Institute for Information \& Communication Technology Planning \& Evaluation) (RS-2024-00347394). (Co-corresponding authors: Myungsuk Kim and Jihong Kim)}
}

\markboth{IEEE Journal of \LaTeX\ Class,~Vol.~12, No.~6, February~2024}%
{Shell \MakeLowercase{\textit{et al.}}: A Sample Article Using IEEEtran.cls for IEEE Journals}

\IEEEaftertitletext{\vspace{-2\baselineskip}}

\maketitle
\begin{abstract}
Although read disturbance has emerged as a major reliability concern, managing read disturbance in modern NAND flash memory has not been thoroughly investigated yet. From a device characterization study using real modern NAND flash memory, we observe that reading a page incurs heterogeneous reliability impacts on each WL, which makes the existing block-level read reclaim extremely inefficient. We propose a new WL-level read-reclaim technique, called \maintech, which keeps track of the accumulated read-disturbance effect on each WL and reclaims only heavily-disturbed WLs. By avoiding unnecessary read-reclaim operations, \maintech reduces read-reclaim-induced page writes by 83.6\% with negligible storage overhead.
\end{abstract}


\section{Introduction}
\label{sec:intro}
NAND flash memory has successfully achieved continuous improvements in storage density over decades, but it has come with significant reliability degradation.
Vertical wordline (WL) stacking and aggressive multi-level cell (MLC) technologies have effectively increased the bit density of flash chips by 2.4$\times$ every two years~\cite{goda2021recent}.
However, such capacity-oriented design decisions 
make modern flash memory significantly more susceptible to various error sources, such as retention loss, read disturbance, and program interference.

In particular, read disturbance has recently gained increasing attention as a major reliability concern in modern (and future) high-density flash memory.
To read a page (from a target wordline (WL)), a flash chip applies a high \emph{pass-through} voltage \vpass (e.g., $>$ 6 V) to all non-target WLs in the same block, which unintentionally programs all non-target WLs slightly and thus can potentially corrupt their stored data when repeated.
As the block size rapidly increases with continuous vertical WL stacking (e.g., a 123-MB block~\cite{isscc-nand-2023}), reading a page disturbs a larger amount of data in the same block.


Despite its importance, how to efficiently manage read disturbance in modern SSDs has yet to be thoroughly investigated; to our knowledge, all prior works on read disturbance in the literature are based on a simple SSD-management task, called \emph{read reclaim (RR)}~\cite{ha2015integrated,zhang2022cocktail}. 
When a block's read count \br (i.e., the number of page reads to the block) exceeds a predefined threshold \bthr, the SSD controller triggers RR to eliminate read-disturbance-induced errors by rewriting (copying) all valid pages in the block to other free pages. 
The additional writes from RR can significantly affect the performance and lifetime of SSDs~\cite{liu2021prolonging}, so it is necessary to
carefully set \bthr for preventing not only read-distrubance-induced data corruption but also unnecessary RR invocations.

In this work, we show that a conventional RR approach causes prohibitive performance overhead to guarantee data reliability in recent high-density 3D flash memory.
In high-density 3D flash memory, reading a page incurs significantly higher disturbance to the target WL's \emph{exact neighbors} compared to the other WLs in the same block~\cite{xiong2018characterizing,ren2023read}.
Such an asymmetry in read disturbance across WLs makes the existing \emph{block-level} RR extremely inefficient.
For example, when pages at the $k$-th WL \wli{k} are read repeatedly, pages at \wli{k-1} and \wli{k+1} may lose their data at a much lower RC value over when pages are randomly accessed over entire WL's. Although the worst-case access pattern (i.e., repeated reads for the same WL) may not be likely in practice, the existing RR approach should handle such a case safely, thus \bthr being set based on the worst-case pattern.  




To mitigate RR overhead, we propose \maintech (\underline{STR}ess-\underline{A}ware \underline{W}L-based read reclaim technique for high-density NAND flash-based SSDs), a new \emph{WL-level} RR technique for modern SSDs which effectively minimizes unnecessary RR at low cost.
The key insight behind \maintech is that read disturbance should be managed at a finer granularity, i.e., per WL, not per block. Since the impact of read disturbance is substantially different depending on the location of a WL, we need a new approach that can selectively reclaim heavily-disturbed WLs only.
To this end, we construct a read-disturbance model that can accurately estimate the impact of a page read on the reliability of each non-target WL in the same block, which enables \maintech to identify any heavily-disturbed WLs and reclaim them in a timely manner. 
For an efficient implementation of \maintech, we leverage the Space-Saving algorithm~\cite{metwally2005efficient} to mitigate the space overhead for tracking the actual read-disturbance effect to WLs in each block.
Our evaluation using the state-of-the-art SSD simulator~\cite{lee2022mqsim} shows that \maintech reduces RR-induced writes by 83.6\% compared to existing RR approaches with negligible space overhead.
\begin{figure}[b]
    \centering
    \vspace{-18pt}
    \includegraphics[width=\linewidth]{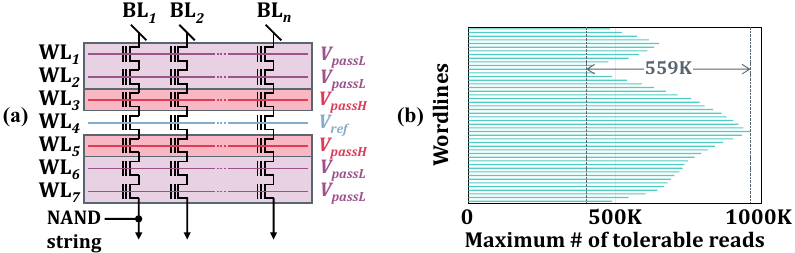}
    \vspace{-20pt}
    \caption{Key factors for asymmetry in read disturbance across WLs.}
    \label{fig:asymmetry}
\end{figure}

\section{Read Disturbance in Modern SSDs}
\label{sec:moti}
\head{Read-Disturbance Variations in Modern Flash Memory}
Unlike in planar (2D) flash memory, where a page read disturbs all non-target WLs in the block almost equally~\cite{ha2015integrated}, the reliability impact of read disturbance significantly varies across WLs in high-density 3D flash memory.
\fig{\ref{fig:asymmetry}} illustrates two key factors contributing to the read-disturbance variations.
\mj{
First, when reading a page, a high-density 3D flash chip applies a higher \vpass (\vpassh, approximately 0.4V higher than \vpassl~\cite{xiong2018characterizing}) to the two adjacent WLs compared to the non-adjacent WLs (\fig{\ref{fig:asymmetry}(a)}). Based on the widely-known Fowler–Nordheim (FN) tunneling equation, the impact of read disturbance is exponentially proportional to \vpass~\cite{ha2015integrated}. Consequently, reading a page leads to a higher reliability impact on the data stored in adjacent WLs~\cite{xiong2018characterizing,ren2023read}. Second, read-disturbance tolerance varies significantly among WLs due to inherent process variations in high-density 3D flash memory~\cite{shim2019exploiting}. \fig{\ref{fig:asymmetry}(b)} shows the maximum tolerable read counts for WLs within a block when the block’s pages are accessed uniformly.
As shown in \fig{\ref{fig:asymmetry}(b)}, the worst WL in a block can tolerate only 403K reads before data corruption, while the best WL can reliably endure 559K additional reads.


To quantify the asymmetry in read disturbance within a high-density 3D block, we measure the raw bit error rate (RBER) of WLs in a block after repeated read operations on a real TLC flash chip. \fig{\ref{fig:moti_disturb_impact}} illustrates the change in RBER for 48 representative WLs in a block under two distinct read patterns, \pa and \pb, where \pa reads only the page at \wli{35}, which is adjacent to the worst WL (\wli{36}) while \pb reads sequentially all pages in the block. As expected, the maximum read count that a block can endure without data loss varies significantly based on the read pattern. \mjr{Under the pattern \pa, reading \wli{35} leads to a significant increase in the RBER of its adjacent WLs (\wli{34} and \wli{36}), resulting in uncorrectable errors (i.e., RBER exceeding the ECC correction capability) after just 54,560 reads. In contrast, the block can tolerate up to 518,420 reliable reads under the pattern \pb.}


\begin{figure}[b]
    \centering
    \vspace{-20pt}
    \includegraphics[width=\linewidth]{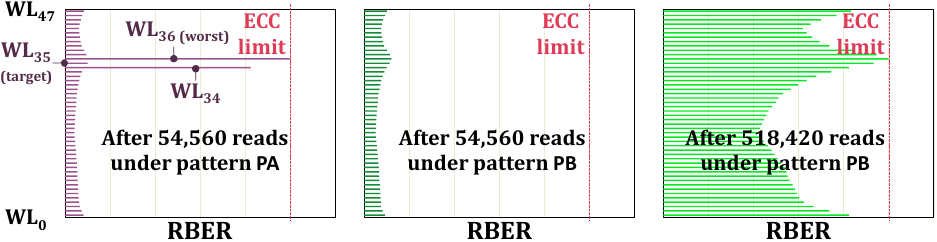}
    \vspace{-20pt}
    \caption{Heterogeneous disturbance impact under different read patterns.}
    \label{fig:moti_disturb_impact}
\end{figure}

\head{Limitations of Existing Solutions}
To evaluate the effectiveness of the conventional RR approach, we measure the number of RR-induced page copies in two SSDs, both employing the block-level RR, \ssdreal and \ssdideal. \ssdreal is an SSD composed of our tested high-density 3D flash memory, while \ssdideal is a hypothetical SSD where read disturbance is symmetric; all non-target WLs in a block experience uniform disturbance, \mjr{with the tolerable read counts of all WLs set equal to the median value shown in }\fig{\ref{fig:asymmetry}(b)}. In both SSDs, \bthr is conservatively set to ensure data reliability even under the worst-case access pattern.

\fig{\ref{fig:moti_overhead}} compares the number of page copies from RR, normalized to \ssdideal, across two distinct P/E cycles under six workloads~\cite{li2023depth}. (For a detailed description of the workloads, see \sect{\ref{sec:eval}}.) 
From the results, we observe that the block-level RR imposes significant lifetime/performance overhead to ensure data reliability in modern high-density 3D flash memory. The number of RR-induced page copies increases by 10.5$\times$, and 15.4$\times$ on average compared to when read disturbance is symmetric, at 1K, and 2K P/E cycles (PEC), respectively. Our evaluation results highlight the fundamental limitations of the block-level RR. 
Due to the heterogeneous reliability impact of read disturbance, data loss can occur at significantly lower \br values under worst-case access patterns (\fig{\ref{fig:moti_disturb_impact}}). Consequently, block-level RR must conservatively set \bthr, leading to frequent and unnecessary RR operations.


\begin{figure}[t]
    \centering
    \includegraphics[width=\linewidth]{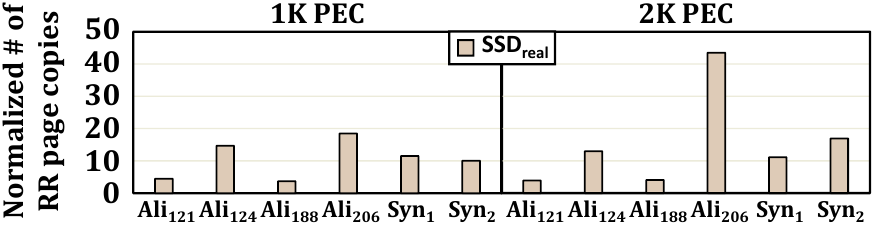}
    \vspace{-20pt}
    \caption{A comparison of the RR overhead between \ssdreal and \ssdideal.}
    \vspace{-15pt}
    \label{fig:moti_overhead}
\end{figure}

}
\section{\js{\maintech: Stress-Aware WL-Based Read Reclaim}}
\label{sec:proposal}
To overcome the limitations of existing solutions, we propose \maintech, a novel WL-level RR technique.
Unlike existing RR techniques that invoke block-level RR based on a conservative \bthr, \maintech reclaims individual WLs only when necessary, thereby significantly reducing the performance and lifetime overheads from RR.
To this end, we develop \inum{i} a \js{new} read-disturbance model that quantifies the heterogeneous reliability impact of read disturbance (\sect{\ref{subsec:proposal_model}}) and \inum{ii} a \maintech-enabled flash translation layer (FTL) that efficiently estimates the actual read disturbance accumulated to each WL (\sect{\ref{subsec:proposal_ftl}}) \mjr{by leveraging an approximate counting algorithm~\cite{metwally2005efficient}.}

\vspace{-5pt}
\subsection{New Read-Disturbance Model}
\label{subsec:proposal_model}
\mj{
\js{We develop a new read-disturbance model through comprehensive characterization of}
160 real 3D TLC flash chips from Samsung. 
\js{Our} proposed model quantifies \js{two} key factors that contribute to the heterogeneous disturbance impact on non-target WLs during a read operation: \inum{i} \js{the} inherent \js{process} variation\js{s across WLs~\cite{shim2019exploiting}} and \inum{ii}\js{ read-disturbance} asymmetry between adjacent and non-adjacent WLs. 
To account for the first factor with minimal overhead, we classify the WLs within each block into four groups, Best, Good, Bad, and Worst, based on their initial RBER values. 
We then measure the maximum read count for each group across 19,200 blocks, while varying PEC and read patterns.}



\begin{figure}[t]
    \centering
    \includegraphics[width=\linewidth]{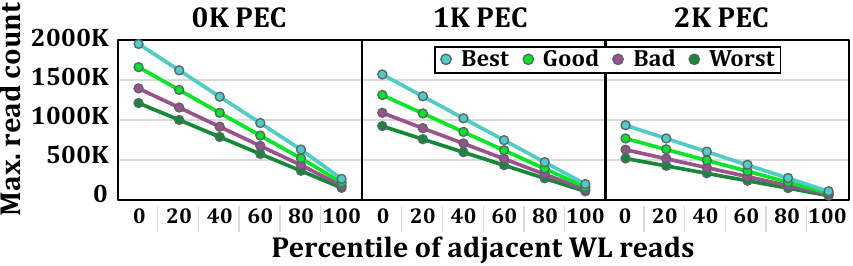}
    \vspace{-22pt}
    \caption{Comparisons of the maximum read counts of each WL Group.}
    \vspace{-10pt}
    \label{fig:device_results}
\end{figure}

\fig{\ref{fig:device_results}} compares the maximum read counts of four WL groups under different PEC and read patterns.
\js{A coordinate ($x$, $y$) for each WL group indicates that the group's worst WL can endure up to $y$ page reads on the same block (i.e., read disturbance from the reads) when $x$\% ($(100-x)$\%) of the reads are performed to (non-)adjacent WLs.}   
We make three key observations. 
First, reading a WL causes significantly more disturbance (8.4$\times$ on average) to adjacent WLs compared to non-adjacent WLs. Second, under the same operating condition, the ratio of read disturbance impact between non-adjacent and adjacent WL reads maintains a consistent disturbance rate $\alpha$. For instance, at 2K PEC, the disturbance rate $\alpha$ is 8.7 for the Best group, indicating that reading the adjacent WL of the worst WL in the Best group (\wlij{w}{Best}) incurs 8.7$\times$ more disturbance stress on \wlij{w}{Best} compared to reading the non-adjacent WL of \wlij{w}{Best}.
Third, the read-disturbance tolerance of a WL group and the disturbance rate $\alpha$ vary significantly depending on operating conditions, such as the inherent reliability characteristics of the WLs and the PEC.

\begin{figure}[t]
    \centering
    \includegraphics[width=\linewidth]{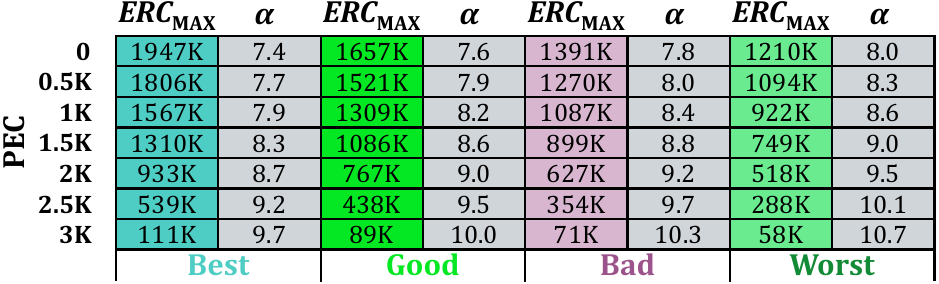}
    \vspace{-18pt}
    \caption{Final Model of \bthr and $\alpha$ under different PEC.}
    \vspace{-15pt}
    \label{fig:model}
\end{figure}

\mjr{
Based on our observations, we derive a model with two key parameters for each WL group: \inum{i} the effective maximum read count, \ercmax, which denotes the maximum number of reads the worst WL in a group can tolerate from non-adjacent WL reads, and \inum{ii} the disturbance rate $\alpha$, which quantifies the relative impact of adjacent WL reads compared to non-adjacent WL reads. The proposed model allows for determining whether a WL is heavily disturbed by using its current effective read count, derived from the read counts of its adjacent and non-adjacent WLs. 
\fig{\ref{fig:model}} shows the parameters of the final model for the tested flash chips under different PEC. For example, at 2K PEC, \wlij{w}{Good} can tolerate 767K non-adjacent reads, and the disturbance rate $\alpha$ is 9.0, respectively.
}


\begin{figure}[b]
    \centering
    \vspace{-12pt}
    \includegraphics[width=\linewidth]{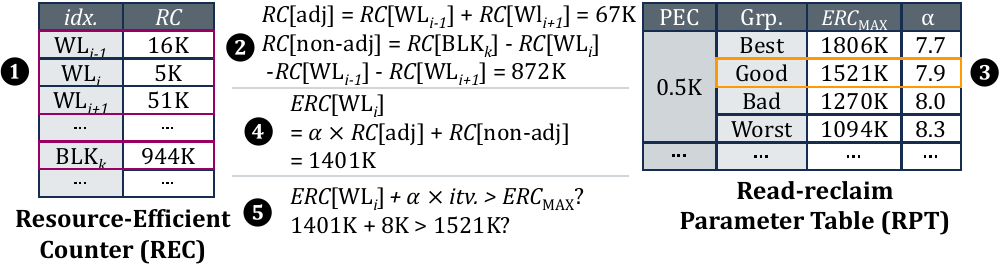}
    \vspace{-20pt}
    \caption{A procedure for identifying heavily-disturbed WLs in \maintech.}
    \label{fig:rr_decision}
\end{figure}

\vspace{-15pt}
\subsection{\ftl}
\label{subsec:proposal_ftl}
We implement an \maintech-enabled FTL, \js{called \ftl,} by extending the conventional page-level FTL~\cite{lee2022mqsim} with two key data structures: \inum{i} \underline{R}ead-reclaim \underline{P}arameter \underline{T}able (RPT) and \inum{ii} \underline{R}esource-\underline{E}fficient \underline{C}ounters (REC). The RPT is a table to store \ercmax and $\alpha$ for each PEC, which can be built through offline profiling of target chips (\fig{\ref{fig:model}}). The REC is a set of per-block counters that keep tracks the \br values of individual WLs within a block, as well as the \br value of the block itself.

\fig{\ref{fig:rr_decision}} illustrates how \ftl estimates the accumulated disturbance impact on individual WLs. For \wli{i}, which is located in the i-th WL in the k-th block, it first looks up the \br values of \wli{i-1}, \wli{i}, \wli{i+1}, and \blki{k} from the REC (\blkcirc{1}). Based on the obtained \br values, \ftl determines the number of reads to adjacent and non-adjacent WLs of \wli{i} (\blkcirc{2}).
Then it queries the RPT with the PEC of \blki{k} and the WL group to which \wli{i} belongs (\blkcirc{3}). \mjr{\ftl converts the number of reads to adjacent and non-adjacent WLs of \wli{i} into the \erc, using the disturbance rate $\alpha$ from the query result (\blkcirc{4}).} The remaining process is straightforward. If the \erc of \wli{i} (including the possible additional reads by the next interval) exceeds \ercmax from the RPT, \ftl identifies \wli{i} as a heavily-disturbed WL (\blkcirc{5}).

\begin{figure}[b]
    \centering
    \vspace{-12pt}
    \includegraphics[width=\linewidth]{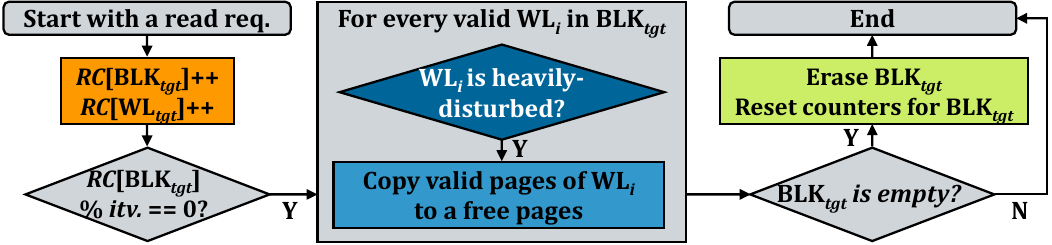}
    \vspace{-15pt}
    \caption{Fine-grained read-disturbance management in \maintech.}
    \label{fig:rr_flow}
\end{figure}

\fig{\ref{fig:rr_flow}} shows how \ftl manages read-disturbance at a WL-granularity. Whenever a page is read, \ftl updates the REC for the target block and WL. Every predefined interval (e.g., every 1K reads to the block), \ftl checks all valid WLs in the block to determine whether the accumulated disturbance impact on any valid WL exceeds the threshold or if there is a possibility it will exceed the threshold by the next interval. For such WLs, \ftl copies the valid pages to free pages before the next interval, thereby preventing read-disturbance-induced data corruption. If the block contains no valid pages after the checking procedure, \ftl erases the block and resets all associated counters.

\head{Overhead Optimization}
\mjr{
\js{\ftl only requires simple modifications over the conventional FTL at high level}, but a naive implementation of per-WL counters \js{introduces non-trivial} storage overhead.
For \js{example, assuming} a 2-TiB SSD comprising a block with 2,568 WLs~\cite{isscc-nand-2023}, the REC requires approximately 125 MB of internal DRAM space (one 3-byte counter per 48-KB WL), which is more than 2,568 times the space required for per-block counting. 
\js{Even though m}odern SSDs typically employ internal DRAM equivalent to 0.1\% of the total SSD capacity, most of this DRAM is dedicated to the address mapping table~\cite{lee2022mqsim}, leaving only a few tens of MB available for other metadata and cache management.

To minimize the storage overhead of per-WL counters,} the REC incorporates the Space-Saving (SS) algorithm~\cite{metwally2005efficient}, which efficiently estimates the frequency of elements in a data stream using a limited number of counters. Each counter entry consists of an element index and its corresponding count value. In our context, the data stream consists of a sequence of reads to a block between successive block erases. 

Upon block erasure, the REC initializes a predefined number of counter entries for that block.
Whenever a WL is read, the REC first checks for an existing counter entry associated with that WL index. If an entry exists, the REC increments the corresponding count value. If no entry exists, the REC identifies the entry with the \emph{lowest} count value, increments its count value, and replaces the element index with the current read WL index.
When the \ftl queries the read count for a particular WL, the REC returns the corresponding count value if an entry exists. If no entry is found, the REC returns the minimum count value among all entries.

Due to the limited number of counters, the estimated read counts by the REC may introduce some error\js{, but} SS ensures that the estimated count value for any element is never underestimated~\cite{metwally2005efficient}. This guarantees that errors in estimation do not result in read-disturbance-induced data corruption, although they may cause premature RR invocations.






\section{Evaluation}
\label{sec:eval}
\head{Evaluation Setup}
\label{subsec:eval_setup}
We evaluate the effectiveness of our proposal using MQSim-E~\cite{lee2022mqsim}, a state-of-the-art SSD simulator. We extend MQSim-E to trigger RR operations according to the read-disturbance characteristics observed in our 19,200 tested blocks.
We configure the architecture and key parameters of the emulated SSDs to closely match those of modern high-performance SSDs employing high-density 3D flash memory. \tab{\ref{tab:eval_setup}} summarizes the simulated SSD configuration. 
We evaluate two synthetic workloads with different I/O patterns, as well as four real-world workloads obtained from Alicloud traces~\cite{li2023depth}. \tab{\ref{tab:eval_workload}} compares the key I/O characteristics of six workloads with varying read ratios and read patterns.

We compare four SSD configurations with different 
RR techniques, \block, \pt, \straw, and \wl. \block is our baseline SSD that employs block-level RR. \pt is an SSD that adopts a state-of-the-art read-disturbance management technique~\cite{han2023page}. Unlike block-level RR, \pt classifies pages within a block according to their page types (e.g., MSB, CSB, and LSB pages in TLC flash memory) and migrates them based on their vulnerability to read disturbance. Both \straw and \wl are SSDs that implement \ftl; however, \straw utilizes the SS algorithms~\cite{metwally2005efficient} for WL-level counters (32 counter entries for a block), while \wl employs naive WL-level counters.

\begin{table}[!b]
    \centering
    \vspace{-16pt}
    \caption{Evaluated SSD configurations.}
    \vspace{-5pt}
    \resizebox{1.0\columnwidth}{!}{%
    \begin{tabular}{l|l}
		\toprule
        \textbf{Configuration} & 2-TiB total capacity; 8 channels; 4 dies/channel; 4 planes/die; \\ & 321 vertical WLs/block; 141 blocks/plane; 7704 pages/block \\
        \midrule
        \textbf{Latencies (\usec)} & \tr $= 40$; \tprog $= 380$; \tbers $= 3500$; \\
        \midrule
        \textbf{Bandwidth} & 8.0 GB/s external I/O bandwidth (PCIe 4.0, 4-lane); \\
                     & 2.0 GB/s channel I/O bandwidth \\
		\bottomrule
    \end{tabular}}
    \label{tab:eval_setup}
\end{table}

\begin{table}[!b]
	\centering
    \vspace{-8pt}
	\caption{Key I/O characteristics of six workloads.}
	\vspace{-5pt}
	\resizebox{1.0\columnwidth}{!}{%
	\begin{tabular}{c|c|c|c|c|c|c}
		\toprule
		\textbf{Workload} & \textbf{Ali$_{121}$} & \textbf{Ali$_{124}$} & \textbf{Ali$_{188}$} & \textbf{Ali$_{206}$} & \textbf{Syn$_{1}$} & \textbf{Syn$_{2}$} \\
		\midrule
		\textbf{Read ratio} & 0.55 & 0.98 & 0.85 & 0.99  & 1.0 & 1.0 \\
		\textbf{Read pattern} & Seq. & Mixed. & Seq. & Rand. & Rand. & Mixed. \\
		\bottomrule
	\end{tabular}}
	\label{tab:eval_workload}%
\end{table}%

\begin{figure}[t]
    \centering
    \includegraphics[width=\linewidth]{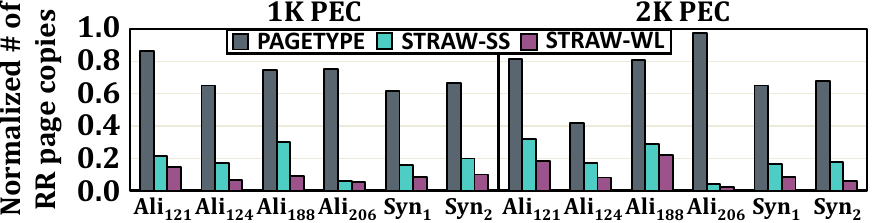}
    \vspace{-20pt}
    \caption{Comparison of RR-induced page copies under two PEC.}
    \label{fig:eval_copies}
\end{figure}

\head{Evaluation Results}
\label{subsec:eval_result}
We first measure the number of pages copies from RR, which is directly related to the effectiveness of read-disturbance management. \fig{\ref{fig:eval_copies}} compares the number of RR-induced page copies in four SSD configurations, normalized to \block, under two different PECs. We make three observations. First, both \straw and \wl significantly reduce the number of RR-induced page copies compared to \block, by preventing premature RR invocations. For example, \straw (\wl) reduces the number of RR-induced page copies over \block by 83.8\% (91.5\%) on average at 2K PEC.
Second, \pt also reduces the number of RR-induced page writes considerably (by
29.4\% on average) compared to \block, but its benefits are limited compared to both \straw and \wl.
Third, with significantly less space overhead, \straw achieves efficiency comparable to \wl under random-read patterns and remains competitive under sequential-read patterns.

\begin{figure}[t]
    \centering
    \vspace{-10pt}
    \includegraphics[width=\linewidth]{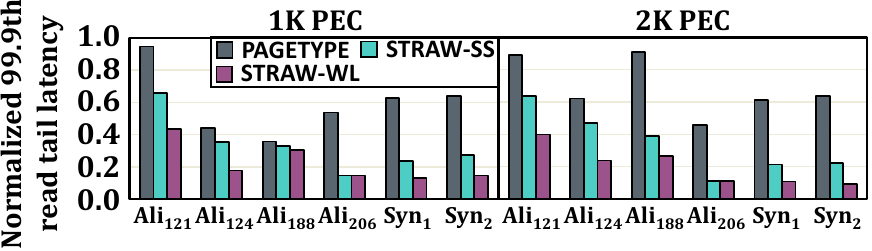}
    \vspace{-20pt}
    \caption{Comparison of 99-th percentile read tail latencies under two PEC.}
    \vspace{-15pt}
    \label{fig:eval_tail}
\end{figure}

We evaluate the impact of reduced RR invocations on read tail latency, which is a crucial performance factor for many data-intensive apps. \fig{\ref{fig:eval_tail}} depicts a comparison of the 99.9th percentile read latencies across the four SSDs at two distinct PEC values. All values are normalized to \block. We make two observations. First, \straw (\wl) significantly reduces the 99.9th percentile read latencies compared to \block by 70.4\% (81.3\%), on average across all the evaluated workloads and PEC. Second, in modern SSDs, the extra operations by RR substantially impact read tail latencies, highlighting the need for efficient read-disturbance management (such as our proposed solution) to meet the strict service level agreements of modern data-intensive apps.

\section{Conclusion}
\label{sec:concl}
We have proposed a new WL-level read reclaim technique, \maintech, which significantly improves SSD lifetime and performance by reducing the frequency of RR invocation. 
Unlike block-level RR that performs RR at block granularity, \maintech identifies heavily-disturbed WLs within blocks and reclaims them in a timely manner. Our evaluation results showed that \maintech effectively enhances SSD lifetime and performance.


\bibliographystyle{IEEEtran}

\begin{thebibliography}{10}

\bibitem{goda2021recent}
A.~Goda.
\newblock {Recent Progress on 3D NAND Flash Technologies}.
\newblock {\em Electronics}, 2021.

\bibitem{isscc-nand-2023}
B.~Kim et~al.
\newblock {28.2 A High-Performance 1Tb 3b/Cell 3D-NAND Flash with a 194MB/s Write Throughput on over 300 Layers $\mathsf{i}$}.
\newblock In {\em ISSCC}, 2023.

\bibitem{ha2015integrated}
K.~Ha et~al.
\newblock {An Integrated Approach for Managing Read Disturbs in High-Density NAND Flash Memory}.
\newblock {\em IEEE TCAD}, 2015.

\bibitem{zhang2022cocktail}
G.~Zhang et~al.
\newblock {Cocktail: Mixing Data with Different Characteristics to Reduce Read Reclaims for NAND Flash Memory}.
\newblock {\em IEEE TCAD}, 2022.

\bibitem{liu2021prolonging}
C.~Liu et~al.
\newblock {Prolonging 3D NAND SSD Lifetime via Read Latency Relaxation}.
\newblock In {\em ASPLOS}, 2021.

\bibitem{xiong2018characterizing}
Q.~Xiong et~al.
\newblock {Characterizing 3D Floating Gate NAND Flash: Observations, Analyses, and Implications}.
\newblock {\em ACM TOS}, 2018.

\bibitem{ren2023read}
T.~Ren et~al.
\newblock {Read Disturb and Reliability: The Complete Story for 3D CT NAND Flash}.
\newblock In {\em NVMSA}, 2023.

\bibitem{metwally2005efficient}
A.~Metwally et~al.
\newblock {Efficient Computation of Frequent and Top-k Elements in Data Streams}.
\newblock In {\em ICDT}, 2005.

\bibitem{lee2022mqsim}
D.~Lee et~al.
\newblock {MQSim-E: An Enterprise SSD Simulator}.
\newblock {\em IEEE CAL}, 2022.

\bibitem{shim2019exploiting}
Y.~Shim et~al.
\newblock {Exploiting Process Similarity of 3D Flash Memory for High Performance SSDs}.
\newblock In {\em MICRO}, 2019.

\bibitem{li2023depth}
J.~Li et~al.
\newblock {An In-Depth Comparative Analysis of Cloud Block Storage Workloads: Findings and Implications}.
\newblock {\em ACM TOS}, 2023.

\bibitem{han2023page}
S.~Han et~al.
\newblock {Page Type-Aware Data Migration Technique for Read Disturb Management of NAND Flash Memory}.
\newblock {\em IEEE TVLSI}, 2023.

\end{thebibliography}

\end{document}